\documentclass[final,1p,times]{elsarticle}

\usepackage{amssymb}
\usepackage{xcolor}
\usepackage{float}
\usepackage{amsmath}

\journal{arXiv}  

\begin{document}

\begin{frontmatter}



\title{Incorrect Resonance Escape Probability in Monte Carlo Codes due to the Threshold Approximation of Temperature-Dependent Scattering}


\author[inst1]{Gabriel Lentchner}

\affiliation[inst1]{organization={The University of Tennessee-Knoxville Department of Nuclear Engineering},
            addressline={863 Neyland Drive}, 
            city={Knoxville},
            postcode={37996}, 
            state={Tennessee},
            country={United States of America}}

\author[inst1]{William Fritsch }
\author[inst1]{Robert Crowder }
\author[inst1]{Noah Walton}
\author[inst1]{Amanda Lewis}
\author[inst1,inst2]{Ondrej Chvala}
\author[inst2]{Kevin Clarno}
\author[inst1]{Vladimir Sobes}

\affiliation[inst2]{organization={The University of Texas-Austen Department of Nuclear Engineering},
            addressline={10100 Burnet Road, PRC 159}, 
            city={Austin},
            postcode={78758}, 
            state={Texas},
            country={United States of America}}

\begin{abstract}
Monte Carlo-transport codes are designed to simulate the complex neutron transport physics associated with nuclear systems. These codes are tasked with simulating phenomena such as temperature effects on cross-sections, thermo-physical effects, reaction rates, and kinematics. It is not computationally possible to simulate the physics of a system exactly. However, many of the approximations made by modern simulation codes have been well validated. This article investigates an impactful simulation error caused by an approximation made in many Monte Carlo-transport codes. The approximation that target-at-rest is valid for neutrons at energies 400 times that of the thermal energy of the target particle is found to be inaccurate in certain scenarios. This paper identifies such cases, notably TRISO \cite{triso_p} fuel and instances where fuel infiltrates the pores of graphite in Molten Salt Reactors. The breakdown of this approximation occurs particularly when there exists a small length scale between fuel, a material with absorption resonances, and moderator, a scattering material. When threshold values are too small, resonance escape probabilities can deviate by as much as 1\% per resonance, forming a baseline defect. Furthermore, two distinct anomalies were observed upon temperature variation, directly attributed to the transition between target-at-rest and target-in-motion physics. Equations provided in this study offer predictions for the temperature ranges within which these anomalies occur, based on system temperature and threshold value. The recommendations put forth in this paper advocate for incorporating the threshold value as a user-defined variable in transport Monte Carlo codes employing this approximation. Additionally, users are advised to conduct convergence studies to ensure that the chosen threshold value is sufficiently high to mitigate the influence of baseline defects and anomalies.
\end{abstract}


\begin{highlights}
\item The threshold between target-at-rest and target-in-motion in most major Monte Carlo transport codes is too small.
\item Setting the threshold too low may lead to inaccuracies in the resonance escape probability of resonances above the threshold, potentially up to 1\% per resonance. 

\item The current threshold approximation does not work, and the anomalies are evident, when there is a small geometric length scale, measured in neutron mean free path, between fuel, material with absorbing resonances, and moderator, scattering material.

\end{highlights}


\end{frontmatter}


\section{Introduction}

In a world where Monte Carlo transport codes are being used daily to license current and new reactors, it is important to always question the results that these codes provide. Built into these transport codes are complex calculations of reactions, scattering kinematics, and temperature effects. All of the complex physics that is present in nuclear systems cannot be modeled exactly and assumptions must be made. This paper examines an approximation long believed to be universally valid for Monte Carlo simulations and concludes that it is not accurate.

Most industry-standard Monte Carlo codes utilize a threshold approximation of temperature-dependent scattering physics. However, setting the threshold too low may lead to inaccuracies in the resonance escape probability for resonances above the threshold, potentially up to 1\% per resonance. This low setting of the threshold is observed across all major Monte Carlo codes employing this implementation. When employing a small threshold value, the approximation that governs the transition between target-at-rest and target-in-motion collision physics results in nonphysical temperature trends and inaccuracies in the determination of resonance escape probability and k$_{eff}$. This effect is visually apparent, especially when isotopes with low-lying resonances, such as $^{238}$U, are combined with moderator materials of intermediate mass, such as Carbon. This erroneous approximation holds significant implications for the nuclear industry in two distinct use cases. Firstly, in TRISO \cite{triso_p} fuel, wherein the proximity of $^{238}$U and Carbon is on a small enough length scale to exhibit this error. Secondly, is a specific condition experienced by Molten Salt Reactors where fuel penetrates the graphite moderator. There may also be other examples where this incorrect approximation is significant. The two cases provided are just the prominent examples which we have identified. This paper will explain the physics of these effects.

It was within this Molten Salt Reactor scenario that this anomaly was initially observed. Abilene Christian University is currently in the process of licensing with the intention to construct a Molten Salt Research Reactor (MSRR) \cite{abilene2022}. Throughout the licensing process of any nuclear reactor in the country, various postulated accidents are modeled to ensure adequate safety. One such event that must be modeled for the MSRR is the penetration of molten fuel from the reactor into the graphite moderator. The infiltration of molten salt into the graphite has been extensively studied and identified as a credible scenario during the Molten Salt Reactor Experiment (MSRE) \cite{msre,porosityInvestigate}. This phenomenon has been attributed to the porosity of graphite, allowing the molten salt fuel to permeate into its pores. It is noteworthy that the graphite moderator of the MSRR can exhibit up to 20\% porosity \cite{CHVALAconfpaper}. The significance lies in the fact that the fuel present in the graphite can influence the reactivity coefficient of the reactor. In the MSRE, this effect resulted in a positive reactivity coefficient owing to the high enriched uranium in the fuel \cite{porosityFeedback}. However, a positive reactivity coefficient is not predicted in the MSRR, as it utilizes low-enriched uranium \cite{CHVALAconfpaper}. Despite the impossibility of a positive reactivity coefficient, it remains crucial to model and comprehend its potential effects on the reactor.

\begin{figure}[H] 
  \centering
  \begin{minipage}[b]{0.35\textwidth} 
    \includegraphics[width=\textwidth]{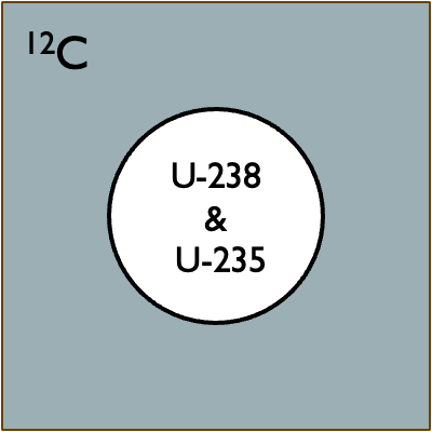}
    \caption{0\% Intrusion simple reflected pin-cell.}
    \label{zero}
  \end{minipage}
  \hfill
  \begin{minipage}[b]{0.35\textwidth} 
    \includegraphics[width=\textwidth]{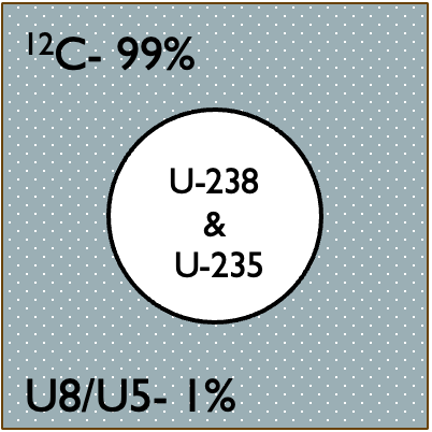}
    \caption{1\% Intrusion simple reflected pin-cell.}
    \label{one}
  \end{minipage}
\end{figure}

Ondrej Chvala, a researcher from the University of Texas-Austin Department of Nuclear Engineering involved in the project, endeavored to quantify the impact of fuel intrusion into graphite on the temperature feedback coefficient \cite{CHVALAconfpaper}. In this pursuit, he modeled the effect using two separate Monte Carlo Transport codes, SCALE 6.3.1 \cite{scale631} and Serpent \cite{serpent}, to compare results \cite{CHVALAconfpaper}. During this cross-comparison process, he detected an anomaly \cite{CHVALAconfpaper}. Even after simplifying the model to a basic reflected pin-cell containing only low-enriched uranium and carbon, the anomaly persisted. For context, Figures~\ref{zero} and \ref{one} illustrate the simple geometry being modeled in both SCALE and Serpent.

\begin{figure}[H]
    \centering
    \includegraphics[width= 13.5cm, height=6cm]{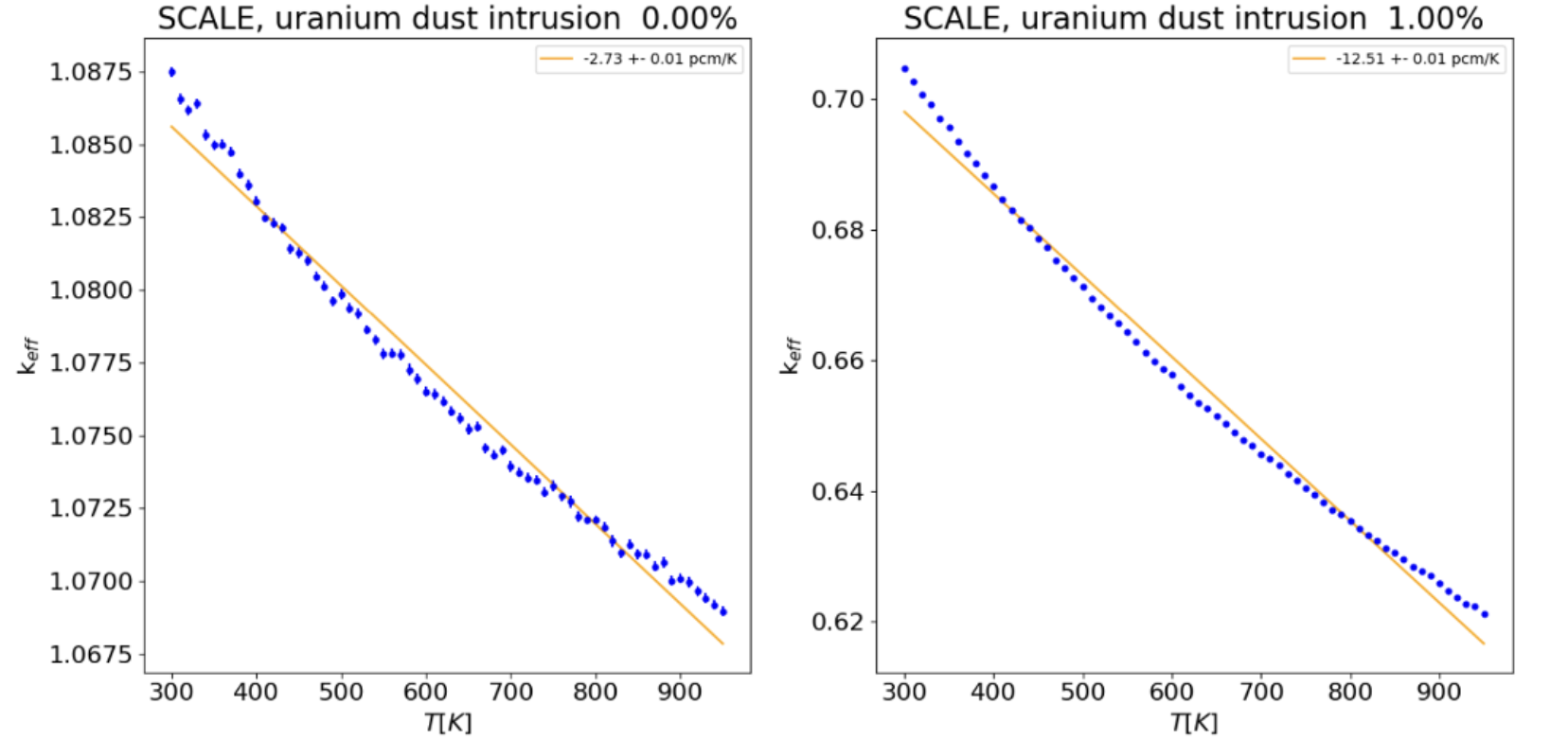}
    \caption{Temperature vs $k_{eff}$ for 0\% and 1\% intrusion simple reflected pin-cell geometry modeled in SCALE 6.3.1 \cite{CHVALAconfpaper}.}
    \label{scaleChvala}
\end{figure}

\begin{figure}[H]
    \centering
    \includegraphics[width= 13.5cm, height=6cm]{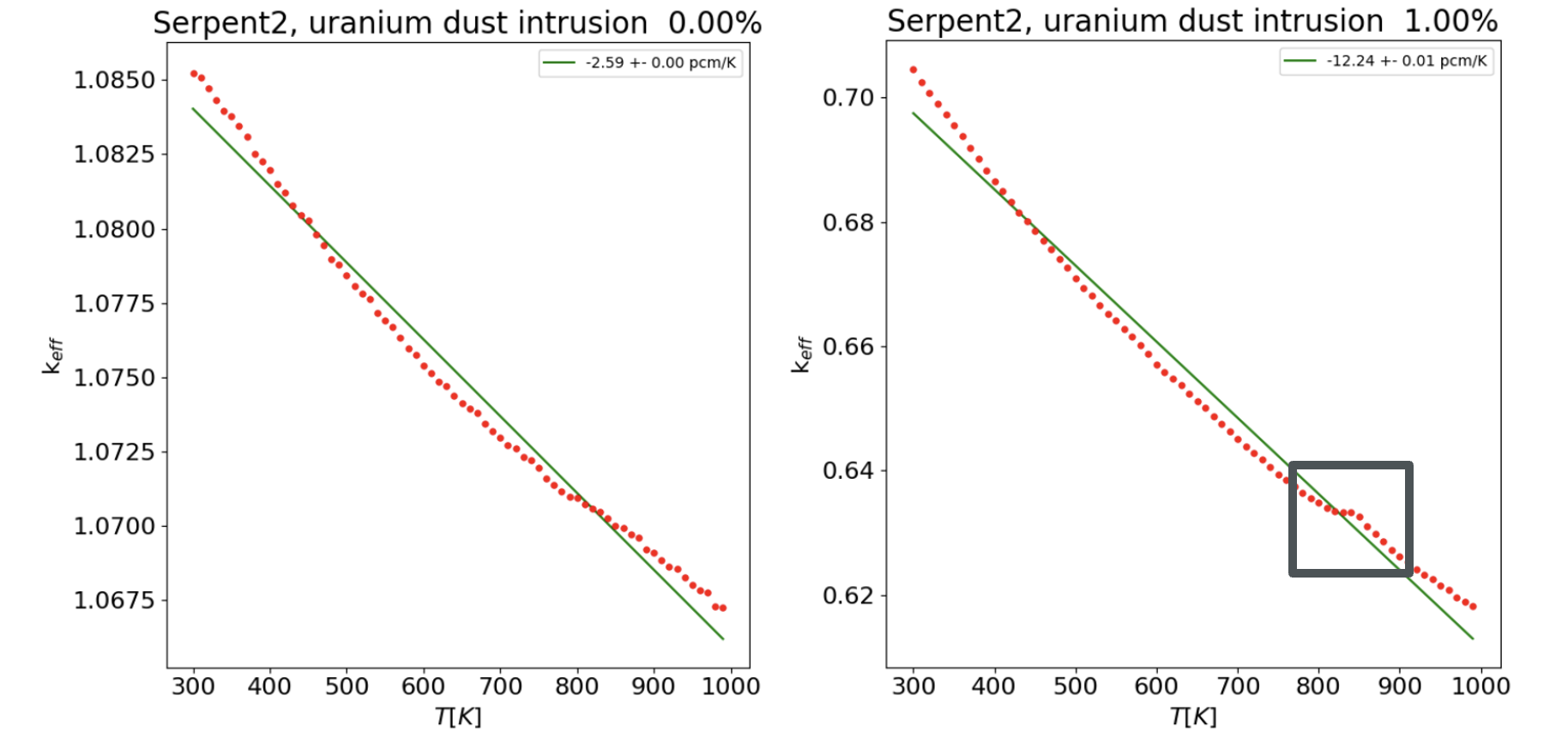}
    \caption{Temperature vs $k_{eff}$ for 0\% and 1\% intrusion simple reflected pin-cell geometry modeled in Serpent \cite{CHVALAconfpaper}.}
    \label{serpentChvala}
\end{figure}

Figure~\ref{scaleChvala} displays the results obtained by Dr. Chvala using SCALE 6.3.1 to simulate the two simple geometries \cite{CHVALAconfpaper}. The leftmost plot in Figure~\ref{scaleChvala} represents the scenario with 0\% intrusion of fuel into the graphite, while the rightmost plot depicts the case with 1\% intrusion of fuel into the graphite. As expected, a nearly linear negative temperature feedback coefficient is observed for this system, evident in both plots of Figure~\ref{scaleChvala}. However, when Serpent was employed to model the same two geometries, an anomaly emerged in the case with 1\% intrusion. This anomaly is illustrated in Figure~\ref{serpentChvala}, highlighted by a gray box surrounding it.

A reactor safety analysis relies on accurate coefficients of temperature-reactivity feedback. In the temperature range of the anomaly, the slope of the reactivity with temperature is wrong. Furthermore, this temperature region is the region of expected operation of graphite moderated high-temperature reactors, making this anomaly highly relevant for safety and licensing of these advanced reactors.

Everyday, advanced nuclear reactors are being developed outside the range of current operational and validated data. Consequently, it becomes vital to cross-compare the established industry codes. The cross-comparison of codes that have been validated against thousands of benchmarks provides confidence in models and simulations that are outside the rigorously tested regimes of these codes. Without cross-comparison, this error may have never been identified. 

This paper will methodically eliminate potential causes of the anomaly until we are certain of the origin of this discrepancy. Then we will describe the error due to the approximation in detail and offer recommendations on how to handle it. 

\section{Theory}

The anomaly arises when the system temperature is varied in a routine temperature feedback coefficient analysis. Only a few physical aspects change with alterations in temperature when neglecting or eliminating thermo-physical properties such as density's variation with temperature. A significant effect of changing temperature involves resonance Doppler broadening. As the temperature increases, the resonances in the system broaden. Doppler broadening of resonances constitutes the primary factor contributing to the negative temperature feedback coefficient. This phenomenon occurs because as the temperature rises, the absorption resonances in the reactor broaden, subsequently decreasing the reactivity of the reactor. Typically, resonance Doppler broadening is handled by processing codes such as NJOY \cite{njoy} or AMPX \cite{ampx}. These processing codes are employed to develop libraries at commonly encountered temperatures for desired isotopes and reactions. Subsequently, these libraries are used by Monte Carlo codes to model particle physics. The process of calculating accurate continuous energy cross-sections at any temperature, reaction, or isotope, is computationally intensive and usually not performed by the Monte Carlo codes themselves. Consequently, temperature libraries were established, commonly available at temperatures of 293.6~K, 600~K, 900~K, 1200~K, and 2500~K. However, the anomaly is not attributed to the implementation of Doppler broadening, as will be demonstrated in the investigation section of this paper.

Another physics calculation that changes in Monte Carlo codes with temperature is Doppler Broadening Rejection Correction (DBRC) \cite{DBRCtRUMBULL,DBRCphdthesis}. DBRC involves a rejection sampling technique used to correct temperature-dependent scattering distributions in regions of highly varying cross-sections, such as resonance peaks. However, it will also be shown that DBRC is not the cause of this effect.

\begin{figure}[H] 
  \centering
  \begin{minipage}[b]{0.45\textwidth} 
    \includegraphics[width=\textwidth]{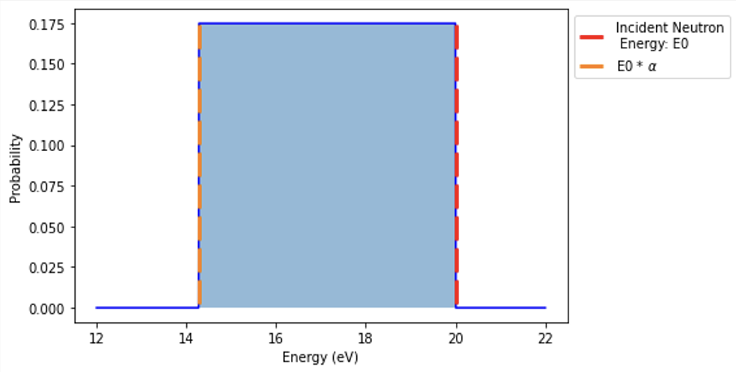}
    \caption{Outgoing neutron energy scattering distribution of off $^{12}$C at 0~K.}
    \label{0k}
  \end{minipage}
  \hfill
  \begin{minipage}[b]{0.45\textwidth} 
    \includegraphics[width=\textwidth]{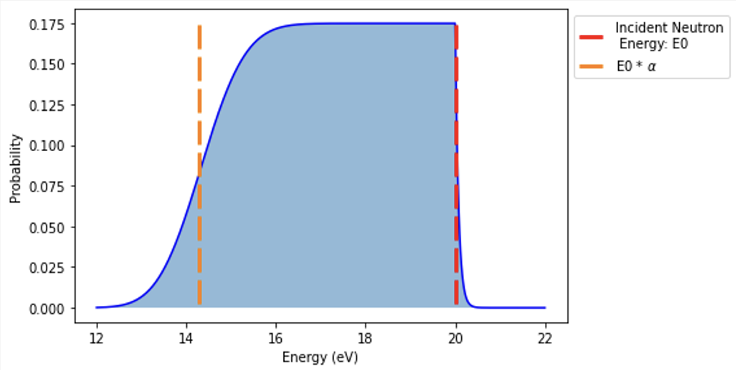}
    \caption{Outgoing neutron energy scattering distribution of off $^{12}$C at 850~K.}
    \label{850k}
  \end{minipage}
\end{figure}

The most critical physics aspect addressed in this paper is the utilization of target-in-motion scattering versus target-at-rest scattering. As the temperature of the system rises, scattering distributions can undergo significant changes, even for isotopes with relatively constant cross-sections. This alteration in scattering distribution is illustrated in Figures~\ref{0k} and \ref{850k}, which depict the difference between the 0~K, constant scattering distribution of outgoing neutron energy, and the 850~K scattering distribution of outgoing neutron energy for $^{12}$C scattering.

It is commonly assumed that target-at-rest scattering physics is applicable when the neutron's energy significantly exceeds the energy of thermal motion for the target particle. However, when the energies are comparable, target-in-motion scattering physics must be employed. Through literature review, it was determined that this approximation is currently set such that target-at-rest scattering physics can be utilized when the neutron energy exceeds 400 times that of the target particle, except for $^1$H, for which target-in-motion scattering is always implemented \cite{DBRCphdthesis,400ktFIRST,400ktMCNP,mcnp5Manual}. This approximation is calculated using Equation~\ref{400kt}, where $E_{th}$ represents the cut-off energy between calculations that consider target-in-motion versus target-at-rest, $k_{B}$ denotes the Boltzmann constant, and $T$ denotes the temperature of the system.

\begin{equation}\label{400kt}
    E_{th}=400\times k_{B}\times T
\end{equation}

The threshold value for the remainder of the paper will be referred to as $\theta$. In Equation~\ref{400kt}, the threshold value, $\theta$ is 400. 

\section{Investigation}

\begin{figure}[H] 
  \centering
  \begin{minipage}[b]{0.49\textwidth} 
    \includegraphics[width=\textwidth]{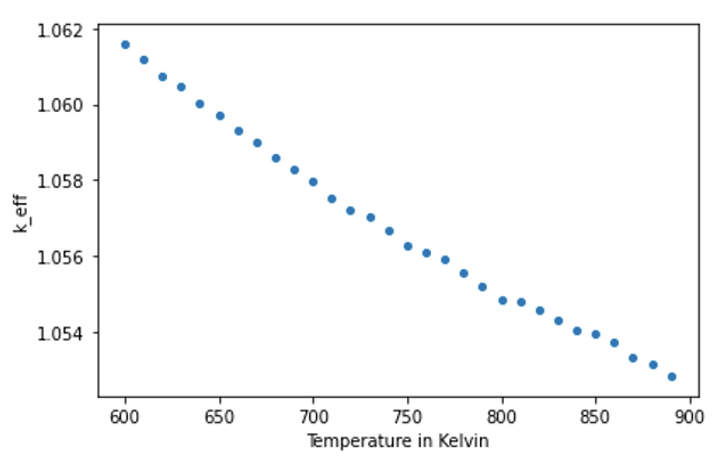}
    \caption{Temperature vs $k_{eff}$ for 0\% intrusion simple reflected pin-cell geometry modeled in MCNP6.2.}
    \label{0int}
  \end{minipage}
  \hfill
  \begin{minipage}[b]{0.49\textwidth} 
    \includegraphics[width=\textwidth]{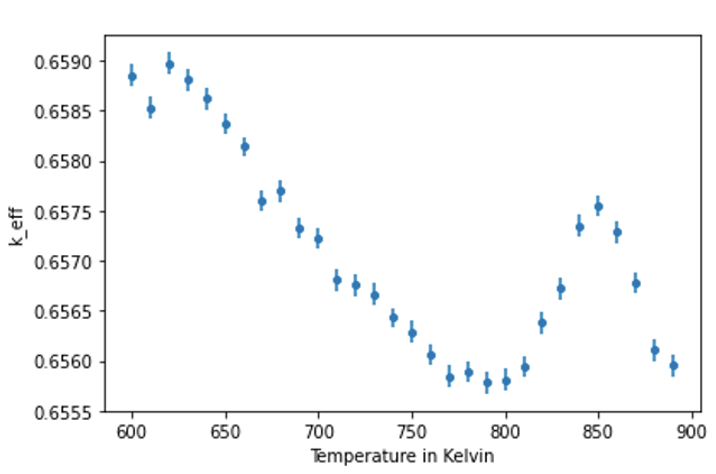}
    \caption{Temperature vs $k_{eff}$ for 1\% intrusion simple reflected pin-cell geometry modeled in MCNP6.2.}
    \label{1int}
  \end{minipage}
\end{figure}

An investigation of the anomaly was conducted to identify and eliminate potential factors contributing to its occurrence, with the aim of narrowing down its cause. This analysis was performed using MCNP 6.2 \cite{mcnp6Manual}. Initially, the geometries modeled by Chvala, et. al. in Serpent and SCALE 6.3 were replicated \cite{CHVALAconfpaper}. Figures~\ref{0int} and \ref{1int} present the results obtained from replicating the same number densities and geometries of the reflected pin-cell models. Figure~\ref{0int} illustrates the case with 0\% intrusion, depicting $k_{eff}$ plotted against the system temperature. On the other hand, Figure~\ref{1int} illustrates the case with 1\% intrusion of fuel, modeled only as $^{235}$U and $^{238}$U, displaying $k_{eff}$ plotted against the system temperature.

Despite having the same geometries and material specifications as the results obtained from SCALE and Serpent, Figures~\ref{0int} and \ref{1int} exhibit significantly different slopes. In addition, the scaling of the plots are different and this is what leads to the anomaly looking more pronounced. This discrepancy arises from the fact that only the 600~K temperature library was provided for the materials in the MCNP run. Unlike SCALE and Serpent, MCNP does not inherently broaden the cross-sections when the ``tmp'' functionality is applied to set the temperature of the cells in the geometry. Consequently, this eliminates the possibility that this effect is caused by the physics of Doppler broadening.

\begin{figure}[H]
    \centering
    \includegraphics[width= 10cm, height=6cm]{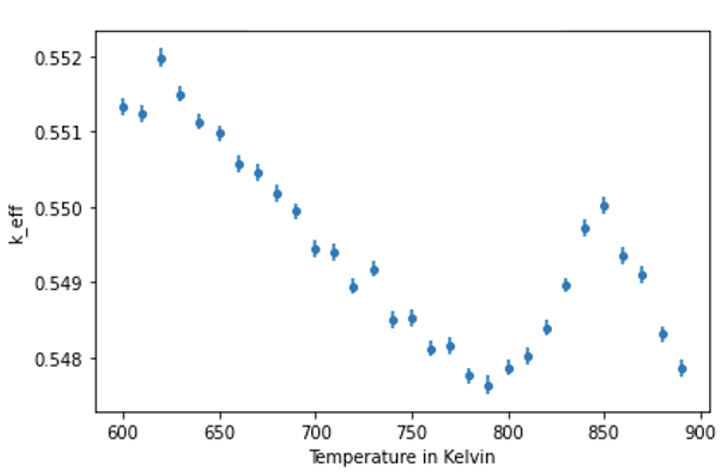}
    \caption{Temperature vs $k_{eff}$ for the homogeneous version of the 0\% intrusion simple reflected pin-cell geometry modeled in MCNP6.2.}
    \label{homo}
\end{figure}

It is imperative to rule out the possibility of a bug in the mixing of materials as a potential cause of the anomaly. A material sampling problem is conceivable because initially, the anomaly was observed when fuel, modeled only as $^{235}$U and $^{238}$U, is mixed with graphite. Figure~\ref{homo} also demonstrates that the anomaly persists in a homogeneous version of the 0\% intrusion case, where temperature vs $k_{eff}$ is still being plotted.

\begin{figure}[H] 
  \centering
    \includegraphics[width=\textwidth]{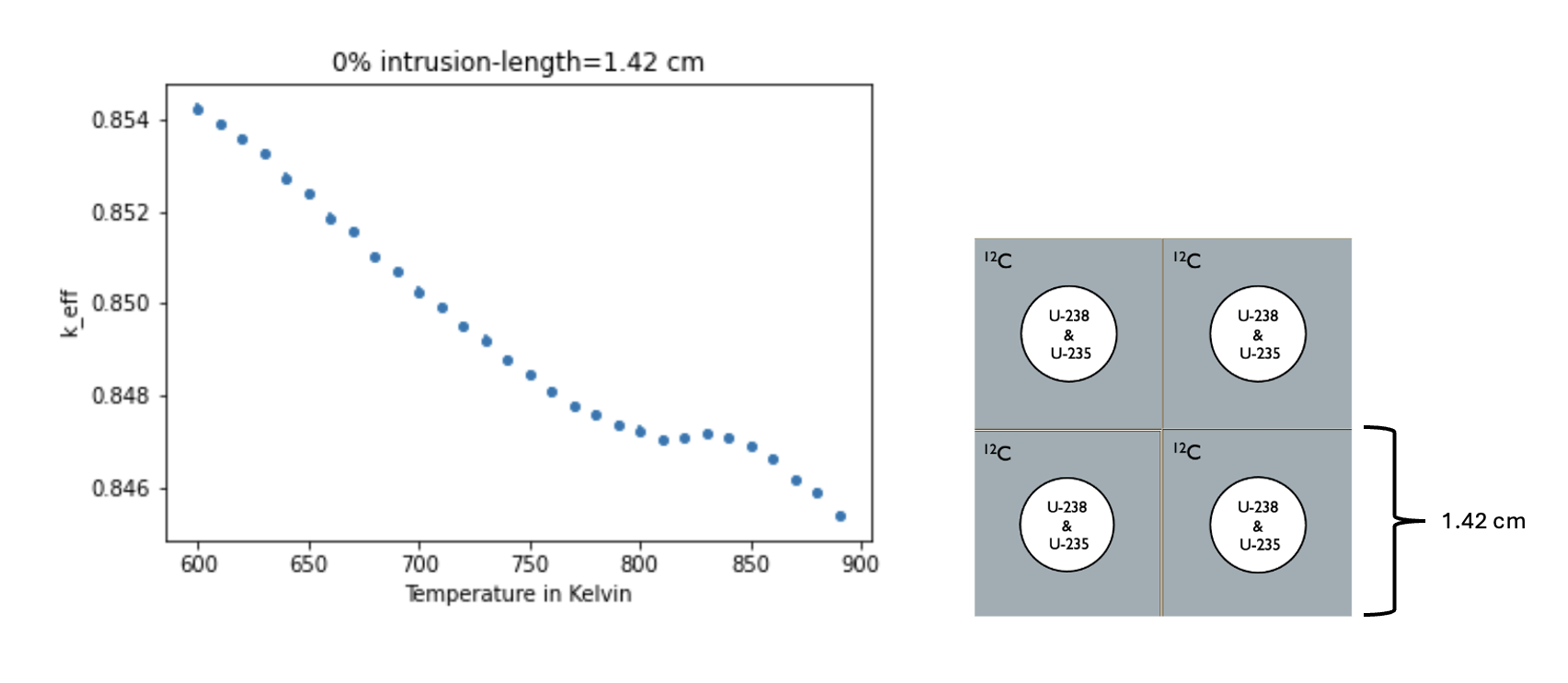}
      \caption{Temperature vs $k_{eff}$ for 0\% intrusion simple reflected pin-cell geometry modeled in MCNP6.2 with the length of the cube modeled was 1.42cm. The diagram on the right is not to scale but represents the effect of shrinking the cube length in reflected boundary conditions. }
    \label{length1_42}
\end{figure}

\begin{figure}[H] 
  \centering
     \includegraphics[width=\textwidth]{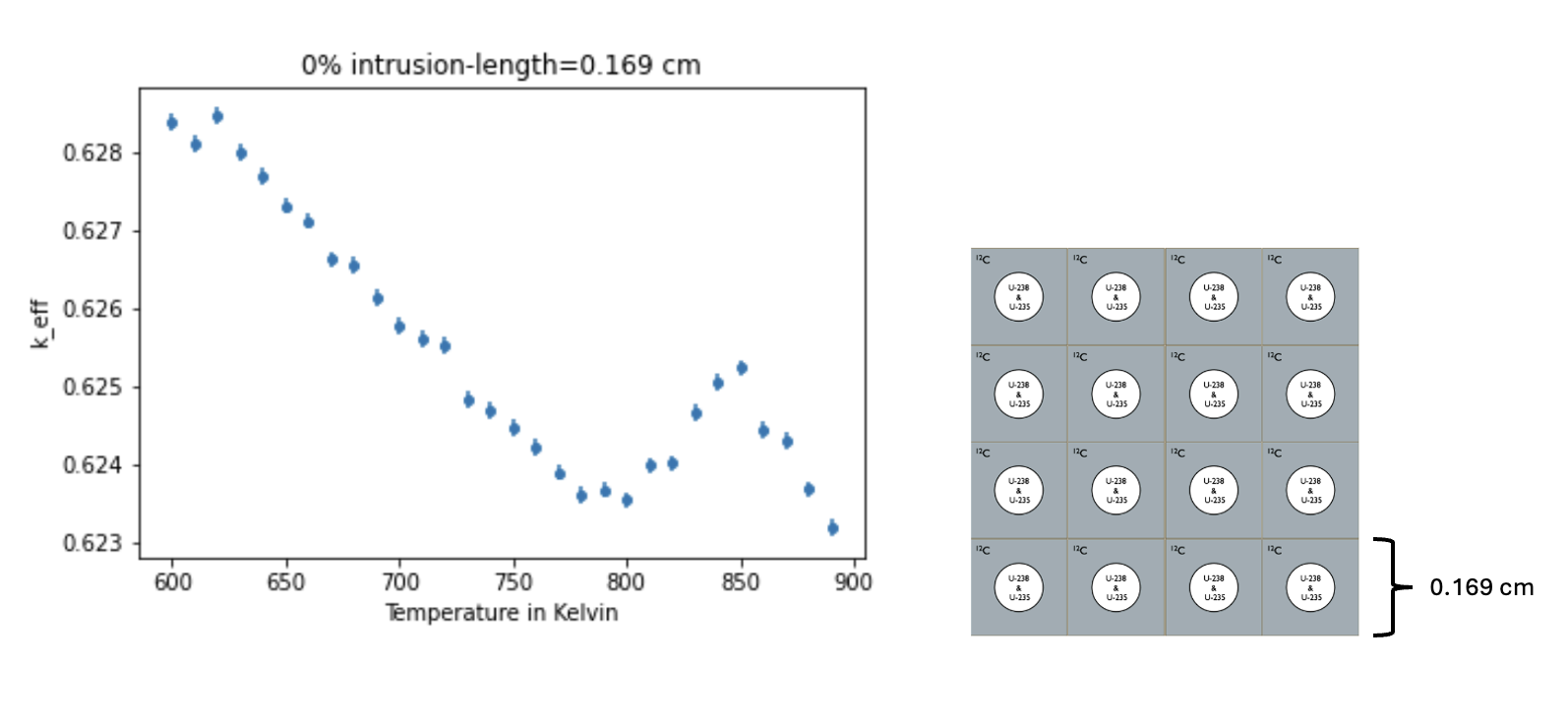}

  \caption{Temperature vs $k_{eff}$ for 0\% intrusion simple reflected pin-cell geometry modeled in MCNP6.2 with the length of the cube modeled was 0.169cm. The diagram on the right is not to scale but represents the effect of shrinking the cube length in reflected boundary conditions.}
    \label{length0_169}
\end{figure}

\begin{figure}[H] 
  \centering
     \includegraphics[width=\textwidth]{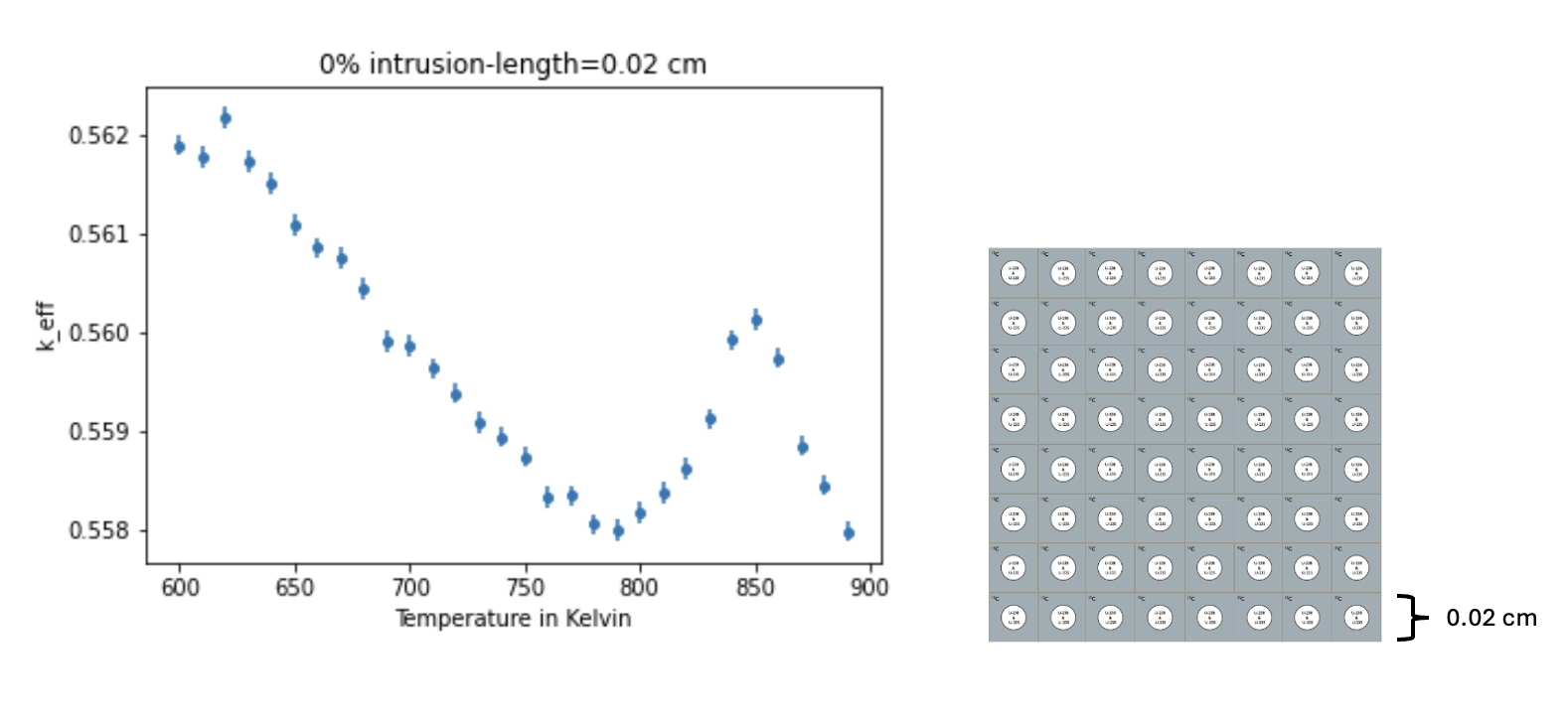}

  \caption{Temperature vs $k_{eff}$ for 0\% intrusion simple reflected pin-cell geometry modeled in MCNP6.2 with the length of the cube modeled was 0.02cm. The diagram on the right is not to scale but represents the effect of shrinking the cube length in reflected boundary conditions.}
    \label{length0_02}
\end{figure}

To demonstrate that the anomaly was not a result of a bug, an approach to homogeneous analysis was undertaken. In neutron transport, reflected heterogeneous geometries converge to the homogeneous solution as the length scale of the reflected pin-cell being modeled approaches zero. To better illustrate this convergence effect, simple diagrams, not to scale, are plotted alongside each of the shrinking length scale plots (Figures~\ref{length1_42}, \ref{length0_169}, and \ref{length0_02}). These diagrams visually depict how the approach to homogeneous process works. The initial length scale for the geometry modeled in Figure~\ref{0int} was 12 cm. A simplified depiction of Figure~\ref{0int} is shown in Figure~\ref{zero}.

Figure~\ref{length1_42} depicts the cube's length reduced to 1.42\,cm, where it can be observed in the temperature vs $k_{eff}$ plot that the anomaly begins to emerge. As the length scale of the model continues to decrease in Figure~\ref{length0_169}, the anomaly progressively grows in size. Finally, in Figure~\ref{length0_02}, the anomaly fully manifests in a model without material mixing, maintaining the same 0\% intrusion model material specifications as Figure~\ref{0int}. In Figure~\ref{length0_02}, the heterogeneous case nearly converges to the solution of the homogeneous case illustrated in Figure~\ref{homo}. Therefore, it can be concluded that the anomaly is not a result of a bug stemming from the sampling of mixed materials. Additionally, it can also be inferred that the anomaly presents itself when the length scale between scattering materials and fuel is very small.

The subsequent step in investigating the anomaly involves determining whether the fuel or the moderating material is the primary cause of the anomaly. This investigation can be conducted using MCNP~6.2, as the ``tmp'' functionality is cell-dependent rather than system-dependent. Consequently, the temperature of the fuel pin can be kept constant while varying the moderator temperature, and vice versa. These scenarios were modeled using a cube length set to 0.169 cm, where the anomaly was observed to occur, for the 0\% intrusion pin-cell model.

\begin{figure}[H]
    \centering
    \includegraphics[width= 10cm, height=6cm]{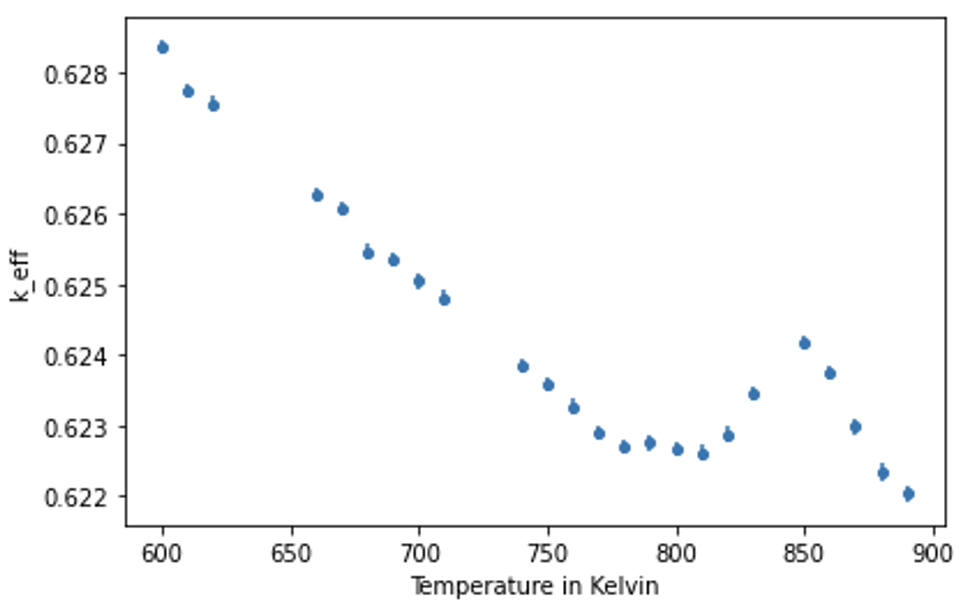}
    \caption{Temperature vs $k_{eff}$ for 0\% intrusion simple reflected pin-cell geometry modeled in MCNP6.2 with the length of the cube modeled being 0.0169cm and the temperature held constant in the fuel pin; the $^{235}$U and $^{238}$U.}
    \label{constantu}
\end{figure}

\begin{figure}[H]
    \centering
    \includegraphics[width= 10cm, height=6cm]{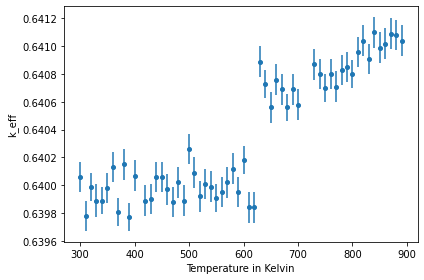}
    \caption{Temperature vs $k_{eff}$ for 0\% intrusion simple reflected pin-cell geometry modeled in MCNP6.2 with the length of the cube modeled being 0.0169cm and the temperature held constant in the moderator; the Carbon.}
    \label{constantc}
\end{figure}

Figure~\ref{constantu} illustrates the temperature vs $k_{eff}$ plot for the 0\% intrusion simple reflected pin-cell geometry modeled in MCNP 6.2, with the cube length set to 0.0169\,cm and the temperature held constant in the fuel pin modeled as only $^{235}$U and $^{238}$U. In this model, the temperature is only varied in the carbon moderator. It can be observed that the anomaly is fully evident in Figure~\ref{constantu}.

Figure~\ref{constantc} depicts the temperature vs $k_{eff}$ plot for the 0\% intrusion simple reflected pin-cell geometry modeled in MCNP 6.2, with the cube length set to 0.0169 cm and the temperature held constant in the carbon surrounding the fuel. In this scenario, the temperature is solely varied in the fuel pin. Notably, Figure~\ref{constantc} does not exhibit the anomaly observed at 850~K. This absence of the anomaly in Figure~\ref{constantc} is significant because it allows for the elimination of DBRC as the cause of the effect. DBRC can be excluded since it is employed to determine scattering distributions accurately in resonance regions, and carbon has a constant cross-section without resonances in this energy range. However, a different, lower anomaly can be seen at 605 K. The lower anomaly will be addressed in detail later in the paper. 

The final simplification undertaken to pinpoint the cause of the anomaly involved converting the k-eigenvalue model into a source-driven problem. Running a source-driven problem allows for the elimination of $^{235}$U from the system to determine if the effect persists. Given that the anomaly has thus far been characterized by an increase in $k_{eff}$, a tally was placed over the low-energy resonances of $^{238}$U, searching for a dip in resonance absorption at 850~K that could potentially contribute to the observed increase.

\begin{figure}[H]
    \centering
    \includegraphics[width= 11cm, height=7cm]{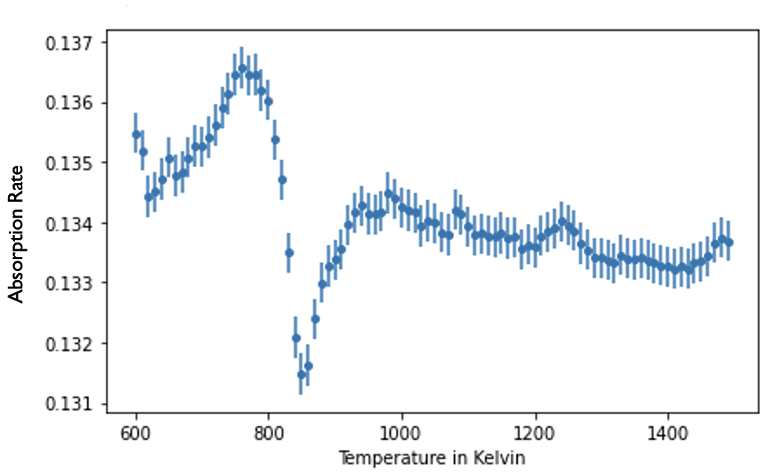}
    \caption{Temperature vs absorption rate over the second resonance, [18,23] eV, of $^{238}$U for homogeneous simple reflected source driven (79 eV source) geometry modeled in MCNP6.2 only of a 1:9 ratio of $^{238}$U to Carbon.}
    \label{homo2ndres}
\end{figure}

\begin{figure}[H]
    \centering
    \includegraphics[width= 11cm, height=6cm]{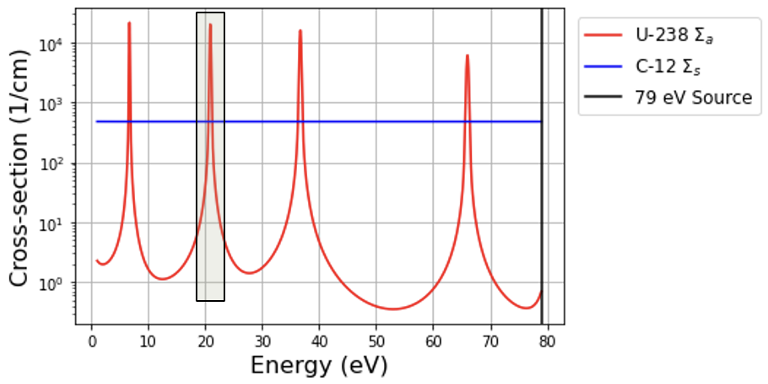}
    \caption{Energy vs macroscopic cross-section with the 79 eV source energy plotted. The shaded black box represents the range that is being tallied over in the data plotted in Figure~\ref{homo2ndres}. These are the macroscopic cross-sections of the system modeled in Figure~\ref{homo2ndres}.}
    \label{u8resonances}
\end{figure}

Figure~\ref{homo2ndres} presents a plot of temperature vs absorption rate over the second resonance of $^{238}$U for a homogeneous simple reflected source-driven geometry modeled in MCNP 6.2, consisting solely of $^{238}$U and carbon. The source in the problem was set at 79\,eV. To provide clarity, Figure~\ref{u8resonances} visually depicts what is being tallied in Figure~\ref{homo2ndres}. The second lowest resonance of $^{238}$U is situated at 20.8\,eV, and Figure~\ref{homo2ndres} displays the absorption rate tallied from the energy range of 18 to 23\,eV, delineated by the gray box highlighted on Figure~\ref{u8resonances}. Clearly visible in Figure~\ref{homo2ndres} is the anomaly manifesting itself in the simplest source-driven problem. Consequently, the anomaly observed at 850~K is attributed solely to the presence of carbon and $^{238}$U.

\section{Results}

\begin{figure}[H]
    \centering
    \includegraphics[width= 15cm, height=6cm]{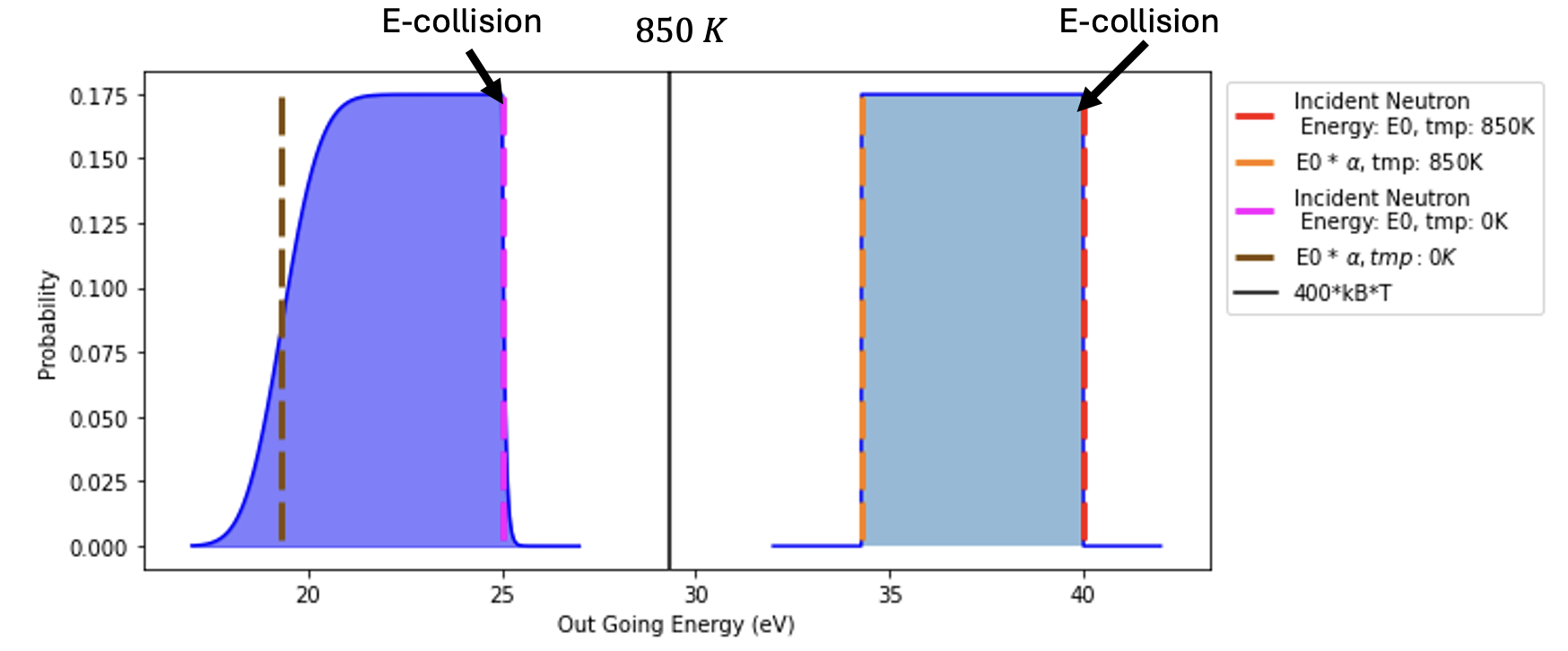}
    \caption{The leftmost distribution represents the scattering distribution when target-in-motion scattering physics is used for a collision energy of 25 eV. The rightmost distribution represents the scattering distribution when target-at-rest scattering physics is applied for a collision energy of 40 eV. The line separating the distributions represents the energy when the switch between methods occurs for the temperature of 850~K. }
    \label{theCause}
\end{figure}

Based on all the information gathered during the investigation of the anomaly, it has been narrowed down to the result of the implementation of target-in-motion scattering versus target-at-rest scattering. The approximation represented by Equation~\ref{400kt} can be visually depicted in Figure~\ref{theCause}. Figure~\ref{theCause} illustrates the disparity between the target-in-motion and target-at-rest scattering distributions. The leftmost distribution represents target-in-motion physics, while the rightmost distribution represents target-at-rest physics. The line delineating these distributions corresponds to the approximation of Equation~\ref{400kt} at 850~K. It is the transition between the implementation of these two physics models that is the root cause of this anomaly.

\begin{figure}[H]
    \centering
    \includegraphics[width=9cm, height=7cm]{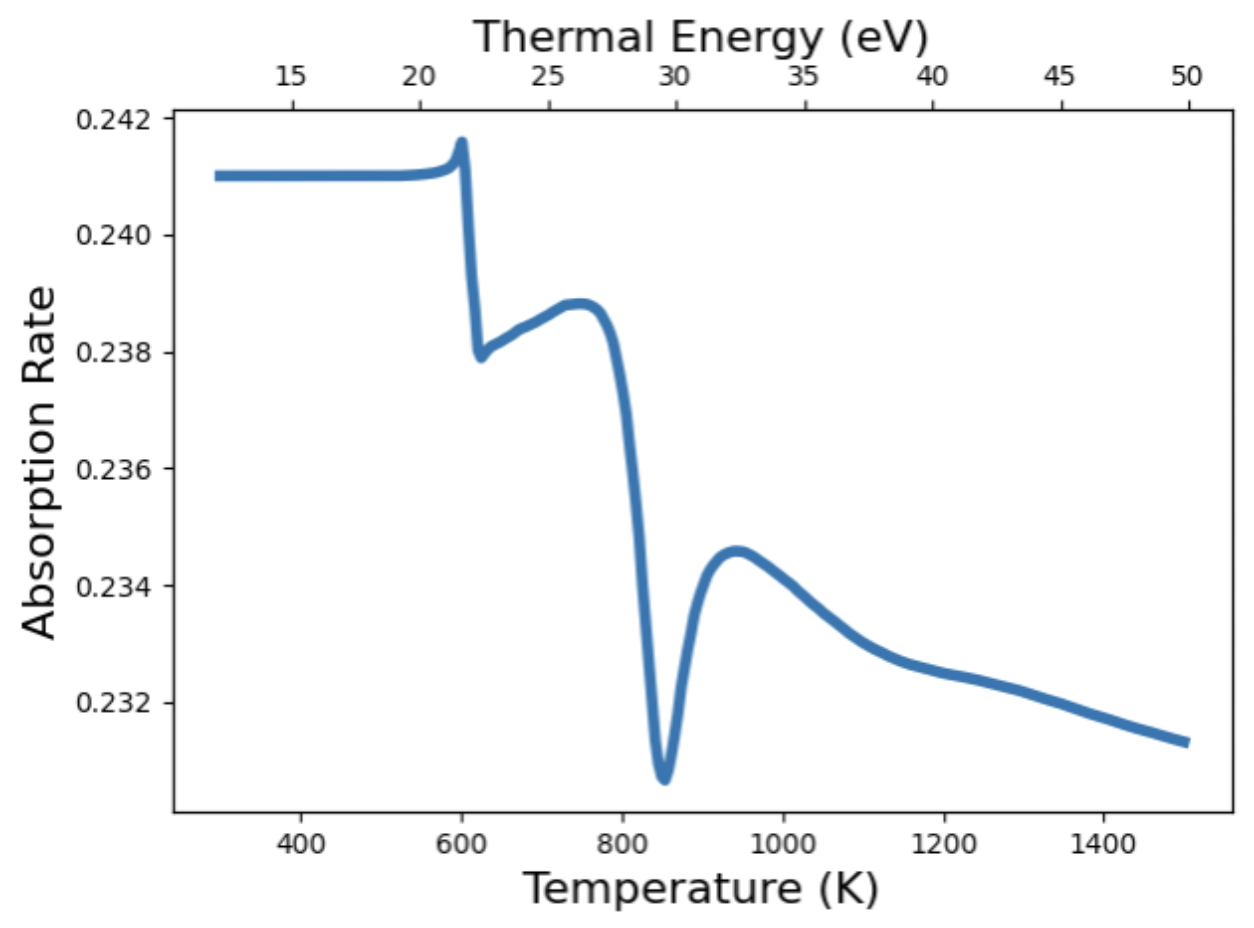}
    \caption{Temperature vs absorption rate over the second resonance, [18,23] eV, of $^{238}$U provided by a deterministic source-driven Python code modelling an infinite homogeneous mixture of $^{238}$U and $^{12}$C. This figure replicates the results of Figure~\ref{homo2ndres}.}
    \label{U8CFirstHumpDeterministic}
\end{figure}

After fully understanding the source of the anomaly and narrowing the problem definition to a simple source-driven problem with only two materials, a source-driven deterministic neutron transport code for infinite homogeneous mixtures was developed in Python. The code, along with its implementation details, is provided in the supplementary material \cite{code_stuff}. The thermal scattering assumption expressed in Equation~\ref{400kt} was replicated within this code. The development of this Python code was informed by relevant research obtained through a literature review. Specifically, the following references were consulted for guidance and insights: \cite{code1, code2, code3}. The deterministic transport code replicates Figure~\ref{homo2ndres} as shown in Figure~\ref{U8CFirstHumpDeterministic}. It can be seen that the anomaly is present in Figure~\ref{U8CFirstHumpDeterministic} at 850~K as expected. Thus, we are confident in the cause of this anomaly.

Armed with this understanding, we were able to develop an equation capable of predicting the location of these anomalies. Equation~\ref{anomaly_e} can accurately predict the precise locations of these humps/dips. In Equation~\ref{anomaly_e}, $E_{\mbox{res}}$ represents the absorption resonance energy of the heavy isotope, $\alpha$ is defined as $\tiny{\frac{\left(\mbox{Moderator Mass - Neutron Mass}\right)^2}{\left(\mbox{Moderator Mass + Neutron Mass}\right)^2}}$, $k_B$ denotes the Boltzmann Constant, and $\theta$ is an arbitrary threshold setting. $\theta$ serves as the multiplicative factor determining the threshold for when target-at-rest scattering should be employed. An infinite value for $\theta$ would imply that target-in-motion physics would be utilized at all neutron energies, with no application of target-at-rest physics.

\begin{equation}\label{anomaly_e}
    T_{\mbox{anomaly}}=\frac{E_{\mbox{res}}/\alpha}{\theta \times k_B}
\end{equation}

Furthermore, Equation~\ref{anomaly_e} highlights several interesting features of the anomaly. Firstly, it reveals that the anomaly occurs when the front edge of the scattering distribution coincides with the resonance while the switch between target-in-motion and target-at-rest scattering occurs. This phenomenon is encapsulated by the $\alpha$ term in Equation~\ref{anomaly_e}. Secondly, Equation~\ref{anomaly_e} suggests that different resonances may experience this dip in resonance absorption. Thirdly, it indicates that different moderators could potentially induce the anomaly at varying temperature locations. Finally, Equation~\ref{anomaly_e} implies that different absorbers may still exhibit the anomaly, and it leads to our recommended solution. These aspects will be explored in detail in the subsequent subsections of the results.

\subsection{Different Resonance}

\begin{figure}[H] 
  \centering

    \includegraphics[width=\textwidth]{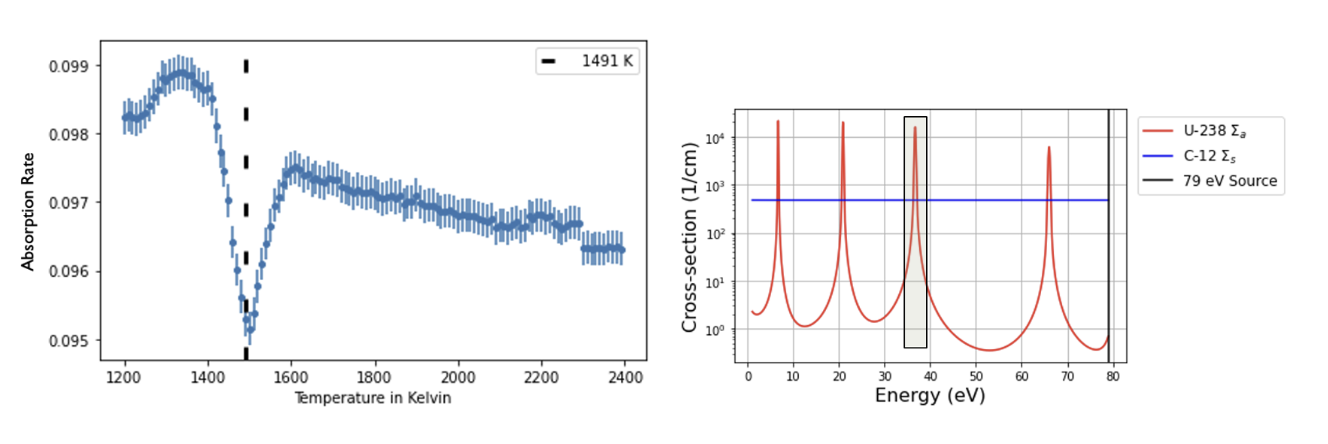}

  \caption{Temperature vs absorption rate over the third resonance, [33,40] eV, of $^{238}$U for homogeneous simple reflected source-driven geometry modeled in MCNP6.2 only of $^{238}$U and Carbon. On the right is the energy vs macroscopic cross-section with the 79\,eV source energy plotted. The shaded black box represents the range that is being tallied over in the data plotted to the left. These are the macroscopic cross-sections of the system modeled in Figure~\ref{homo2ndres}.}
    \label{u83rdres}
\end{figure}

Equation~\ref{anomaly_e} exhibits no limitations; the dip in resonance absorption should be observable across other resonances as well. The leftmost image in Figure~\ref{u83rdres} demonstrates the anomaly across the third resonance of $^{238}$U, while the rightmost figure highlights the region over which the absorption rate is tallied. Equation~\ref{cu3} represents the solution to Equation~\ref{anomaly_e} for the moderator material being carbon and the third resonance of $^{238}$U. Remarkably, Equation~\ref{cu3} accurately predicts the temperature location of the dip in resonance absorption.

\begin{equation}\label{cu3}
     1491 K=\frac{{36.7[eV]}/0.71}{400 \times 8.61\times 10^{-5}[eV*K^{-1}]}
\end{equation}

\subsection{Different Moderator}

\begin{figure}[H] 
  \centering
  \begin{minipage}[b]{0.49\textwidth} 
    \includegraphics[width=\textwidth]{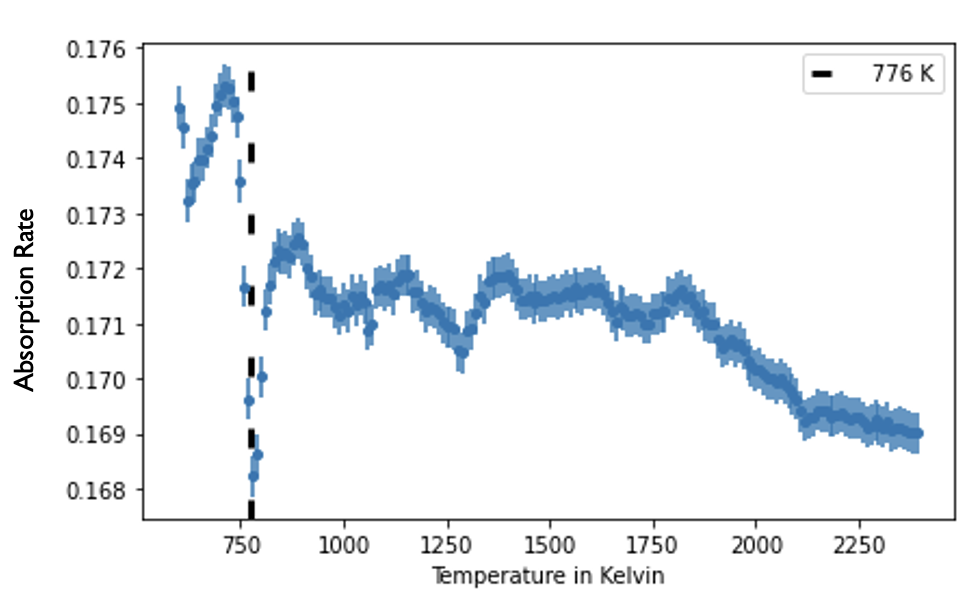}
    \caption{Temperature vs absorption rate over the second resonance, [18,23] eV, of $^{238}$U for homogeneous simple reflected source driven, 79\,eV source, geometry modeled in MCNP6.2 only of a ratio of 1:9 of $^{238}$U to $^{16}$O.}
    \label{U8O2ndres}
  \end{minipage}
  \hfill
  \begin{minipage}[b]{0.49\textwidth} 
    \includegraphics[width=\textwidth]{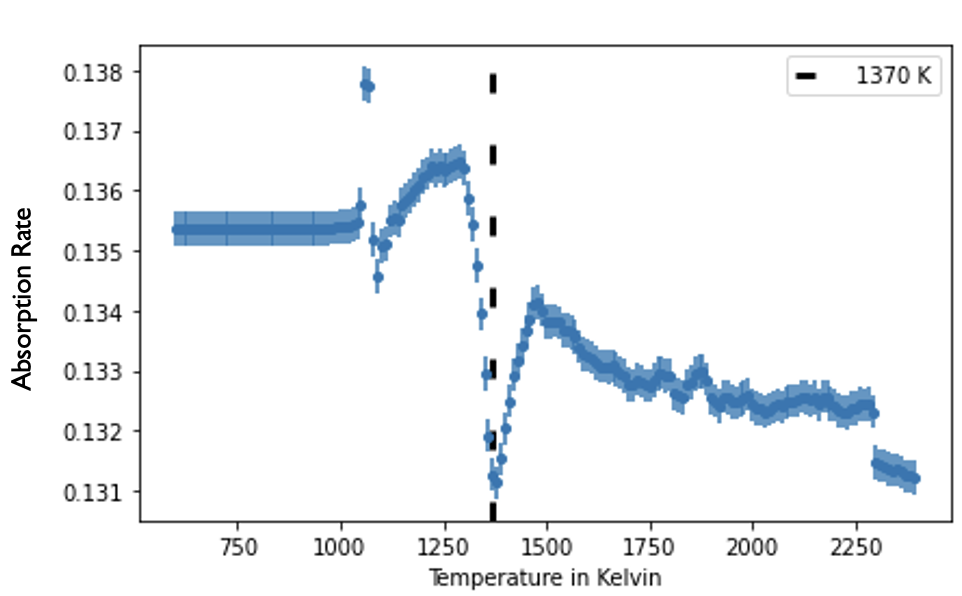}
    \caption{Temperature vs absorption rate over the third resonance, [33,40] eV, of $^{238}$U for homogeneous simple reflected source driven, 79\,eV source, geometry modeled in MCNP6.2 only of a ratio of 1:9 of $^{238}$U to $^{16}$O.}
    \label{u8O3rdres}
  \end{minipage}
\end{figure}

Moreover, owing to the generality of Equation~\ref{anomaly_e}, the dip in resonance absorption should be observable when other ``moderators'', materials with relatively constant scattering cross-sections are used. Figures~\ref{U8O2ndres} and \ref{u8O3rdres} are modeled similarly to Figure~\ref{homo2ndres} (a 79\,eV source-driven, homogeneous problem with reflected boundaries), except that the carbon in the material card was replaced with $^{16}$O. Equation~\ref{ou2} represents the solution to Equation~\ref{anomaly_e} for the moderator material being $^{16}$O and the second resonance of $^{238}$U. Similarly, Equation~\ref{ou3} represents the solution to Equation~\ref{anomaly_e} for the moderator material being $^{16}$O and the third resonance of $^{238}$U. Notably, Equation~\ref{anomaly_e} accurately predicts the location of the dip in resonance absorption for both resonances of $^{238}$U with $^{16}$O as the moderator material.

\begin{equation}\label{ou2}
     776 K=\frac{{20.8[eV]}/0.776}{400 \times 8.61\times 10^{-5}[eV*K^{-1}]}  
\end{equation}
\begin{equation}\label{ou3}
1370 K=\frac{{36.7[eV]}/0.776}{400 \times 8.61\times 10^{-5}[eV*K^{-1}]}
\end{equation}

\begin{figure}[H]
    \centering
    \includegraphics[width= 15cm, height=6cm]{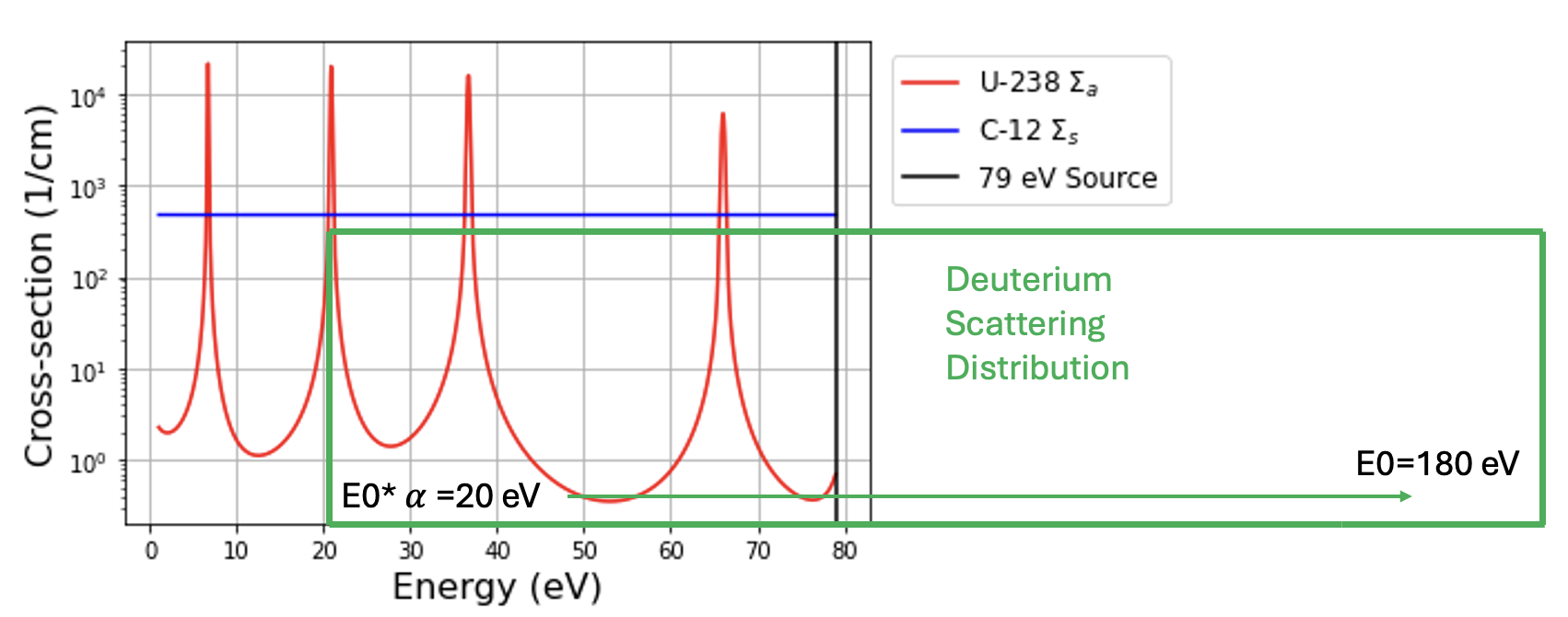}
    \caption{Energy vs macroscopic cross-section with the 79\,eV source energy plotted. The green box represents the deuterium scattering distribution.}
    \label{deuteriumDiagram}
\end{figure}

However, there are instances where the anomaly becomes undetectable. Specifically, if the scattering distribution spans across multiple resonances, the dip in resonance interference is not discernible, nor is a bump in $k_{eff}$ observed. An example of a moderator with a sufficiently wide scattering distribution to experience multi-resonance interference is deuterium ($^2$H). Figure~\ref{deuteriumDiagram} presents a plot of energy vs macroscopic cross-section with a 79\,eV source, alongside a green box representing the deuterium scattering distribution. The front edge of the deuterium scattering distribution is situated within the second resonance of $^{238}$U. In this case, the same as Figure~\ref{homo2ndres} except with deuterium as the moderator, no anomaly is found because the scattering distribution stretches over additional $^{238}$U resonances. This multi-resonance interference obscures the anomaly, rendering it invisible.

\subsection{Different Absorber}

\begin{figure}[H]
    \centering
    \includegraphics[width= 11cm, height=7cm]{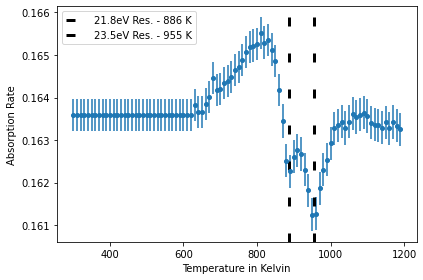}
    \caption{Temperature vs absorption rate tallied over, [21,25] eV, two low resonances of $^{232}$Th (21.8\,eV and 23.5\,eV shown in Figure \ref{thorium_res}) for homogeneous simple reflected source driven, 79\,eV source, geometry modeled in MCNP6.2 only of a ratio of 1:9 of $^{232}$Th to Carbon.}
    \label{thorium_carbon}
\end{figure}

\begin{figure}[H]
    \centering
    \includegraphics[width= 11cm, height=7cm]{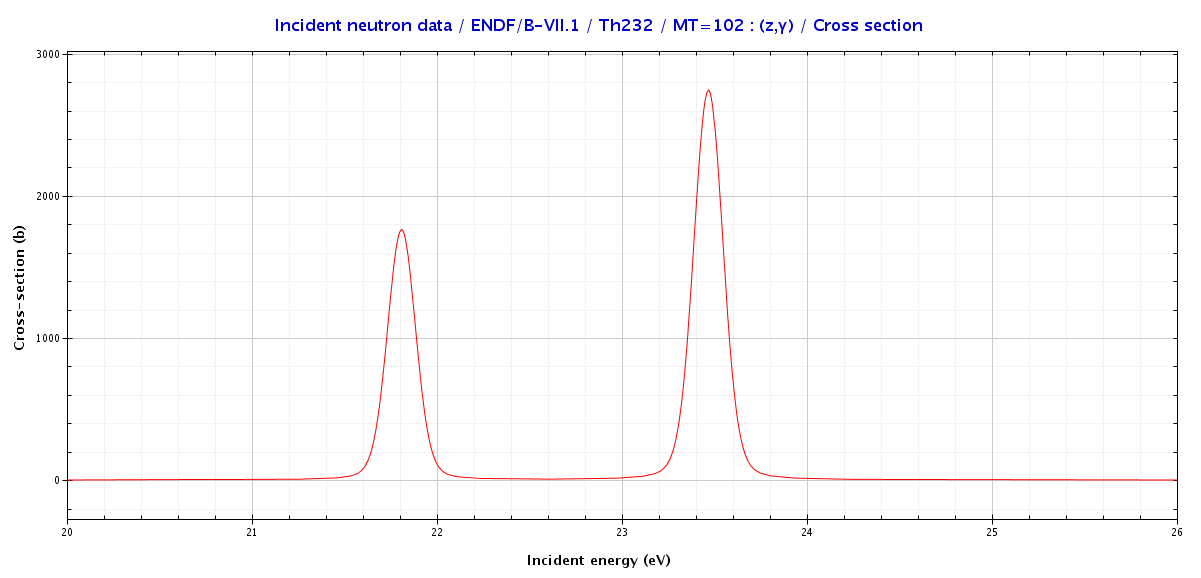}
    \caption{The ENDF/B-VII.I $^{232}$Th radiative capture cross-section showing the two resonances of interest in Figure \ref{thorium_carbon}. This figure was generated using JANIS \cite{janis}.}
    \label{thorium_res}
\end{figure}

Equation~\ref{anomaly_e} also enables exploration of scenarios involving different absorbing materials. Any material with a low-lying absorption resonance, such as $^{238}$U, could potentially exhibit the characteristic dip in resonance absorption. Figure~\ref{thorium_carbon} illustrates this phenomenon with $^{232}$Th. Modeled similarly to Figure~\ref{homo2ndres} (a 79\,eV source-driven, homogeneous problem with reflected boundaries), Figure~\ref{thorium_carbon} replaces the $^{238}$U in the material card with $^{232}$Th. Equation~\ref{anomaly_e} accurately predicts the location of the dip in resonance absorption for both resonances of $^{232}$Th with carbon as the moderator material, occurring at 886~K and 955~K. The absorption, radiative capture, resonances of $^{232}$Th that create the two dips can be seen in Figure \ref{thorium_res}, which was generated using JANIS \cite{janis}. Figure~\ref{thorium_carbon} clearly depicts the dip in resonance absorption precisely at 886~K and 955~K.

\subsection{Secondary Anomaly}

\begin{figure}[H]
    \centering
    \includegraphics[width= 11cm, height=7cm]{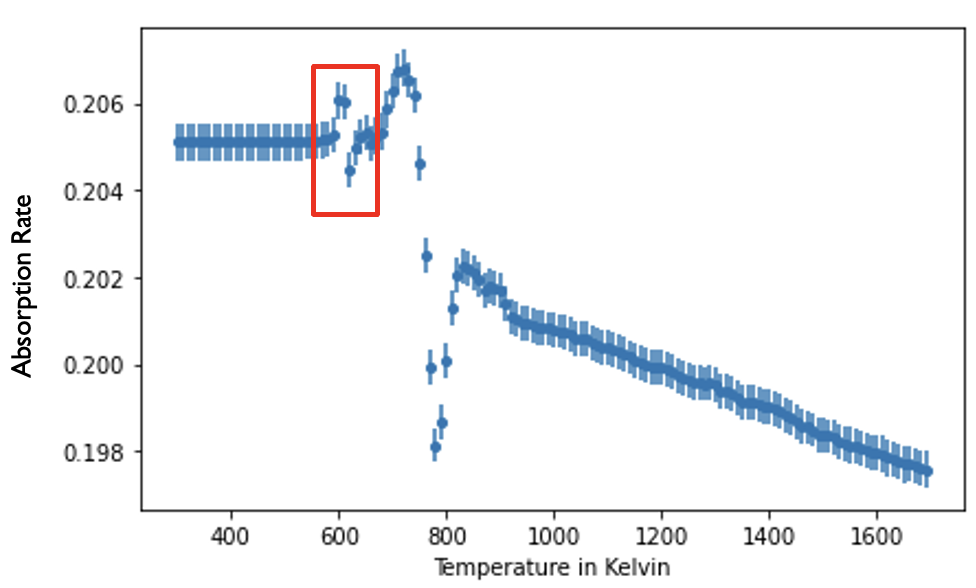}
    \caption{Temperature vs absorption rate over the second resonance, [18,23] eV, of $^{238}$U for homogeneous simple reflected source driven, 32\,eV source, geometry modeled in MCNP6.2 only of $^{238}$U and $^{16}$O. The red box highlights the second anomaly.}
    \label{U8CLowerHump}
\end{figure}

During the analysis of the initial anomaly, a separate secondary anomaly was identified. This second anomaly can be observed in Figure~\ref{U8CLowerHump}, outlined by the red box. Figure~\ref{U8CLowerHump} depicts the same homogeneous reflected geometry with $^{238}$U and $^{16}$O, but the temperature axis now ranges from 300 to 1600~K. When the temperature range is expanded, another nonphysical effect becomes apparent. Through the process of elimination that has already been conducted, it is determined that the approximation of target-in-motion versus target-at-rest scattering is the underlying cause. However, this secondary anomaly is distinct from the moderator and is solely attributable to the absorbing resonances of the ``fuel" material. The secondary anomaly arises when the change in physics occurs within the resonance of the absorber.

Drawing upon this understanding, it becomes feasible to formulate another equation to predict this secondary anomaly. Equation~\ref{second anomaly} is capable of predicting the secondary anomaly, where $E_{\mbox{res}}$ represents the absorption resonance energy of the heavy isotope, $k_B$ denotes the Boltzmann Constant, and $\theta$ is an arbitrary setting. Comparing Equation~\ref{second anomaly} with Equation~\ref{anomaly_e} reveals the difference between the two anomalies. The only disparity between the equations lies in the absence of $\alpha$ in the second anomaly equation. Consequently, the second anomaly exhibits no dependence on the scattering/moderator material.

\begin{equation}\label{second anomaly}
    T_{\mbox{second anomaly}}= \frac{E_{\mbox{res}}}{\theta \times k_B}
\end{equation}

\begin{figure}[H]
    \centering
    \includegraphics[width= 9cm, height=7cm]{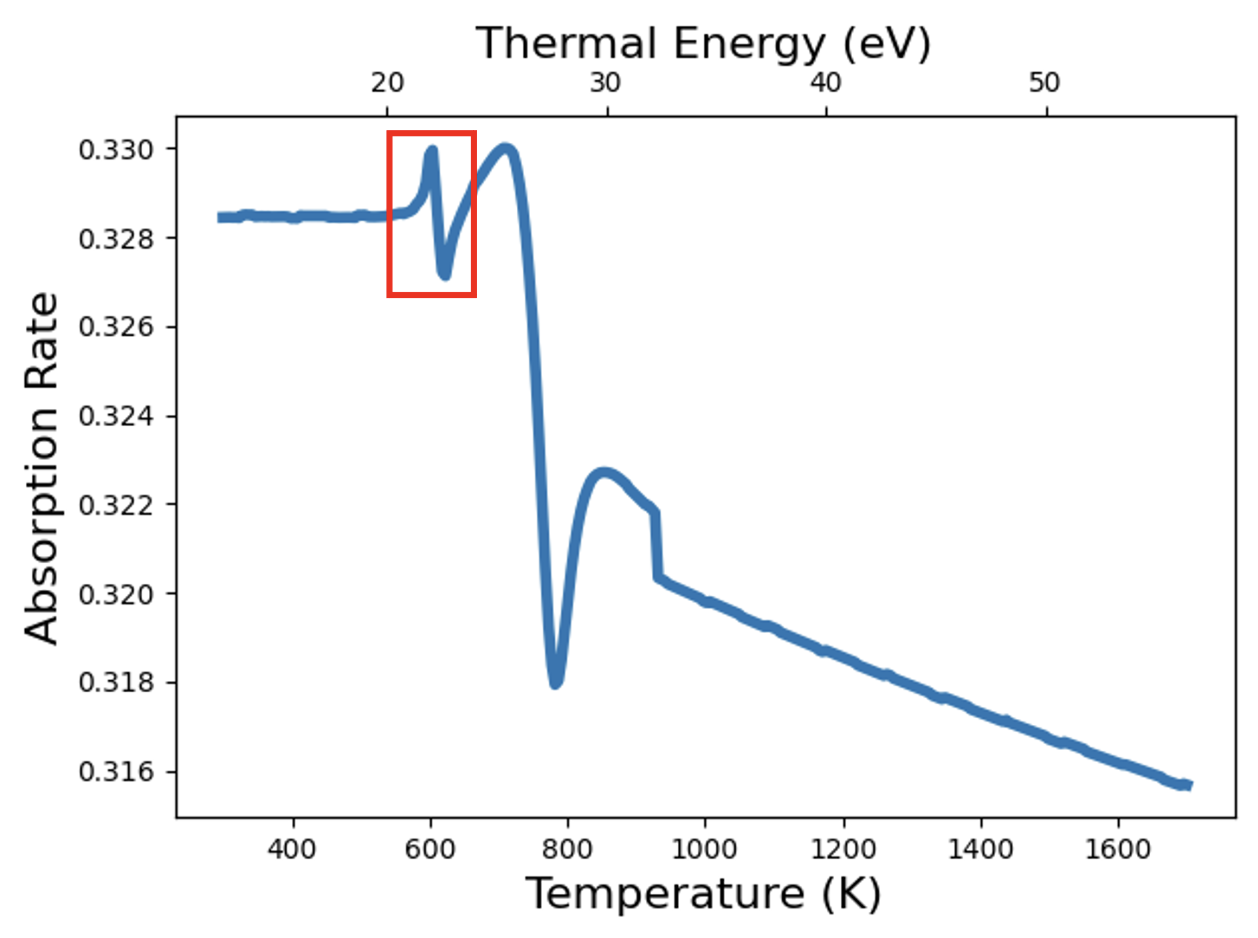}
    \caption{Temperature vs absorption rate over the second resonance, [18,23] eV, of $^{238}$U from a 0-dimensional deterministic source-driven transport code in written in Python. Neutrons are born at 32\,eV and terminated at 12\,eV. The geometry only includes $^{238}$U scattering and capture and  $^{16}$O scattering. The red box highlights the second anomaly.}
    \label{U8CLowerbump_deterministic}
\end{figure}

After understanding the source of the second anomaly and narrowing the problem definition to a simple source-driven problem with only two materials, a simple deterministic code written in Python can replicate this effect. Figure~\ref{U8CLowerbump_deterministic} is the simple Python deterministic code solution to Figure~\ref{U8CLowerHump}. It can be seen that the secondary anomaly is present in Figure~\ref{U8CLowerbump_deterministic}; highlighted by the red box. 

Since Equation~\ref{second anomaly} is also not specific, all the additional cases that worked for Equation~\ref{anomaly_e} will exhibit the hump. To illustrate this, Equation~\ref{second anomaly} is solved in Equation~\ref{ou3_2} for the third resonance of $^{238}$U with $^{16}$O as the moderator material. The calculation predicts a secondary anomaly at 1064 K. Figure~\ref{U8O2ndres} clearly shows the secondary anomaly at that temperature point.

\begin{equation}\label{ou3_2}
1064 K=\frac{{36.7[eV]}}{400 \times 8.61\times 10^{-5}[eV*K^{-1}]}
\end{equation}

However, there is one additional difference between the first and second anomalies. The second anomaly does not suffer from the issue of multi-resonance interference. This is because the second anomaly does not depend on the moderator. Therefore, it remains unaffected by the scattering distribution of the moderator material stretching over multiple resonances. The second anomaly will consistently manifest wherever the transition in physics occurs within a resonance.

\section{Consequences}

\begin{figure}[H]
    \centering
    \includegraphics[width= 11cm, height=7cm]{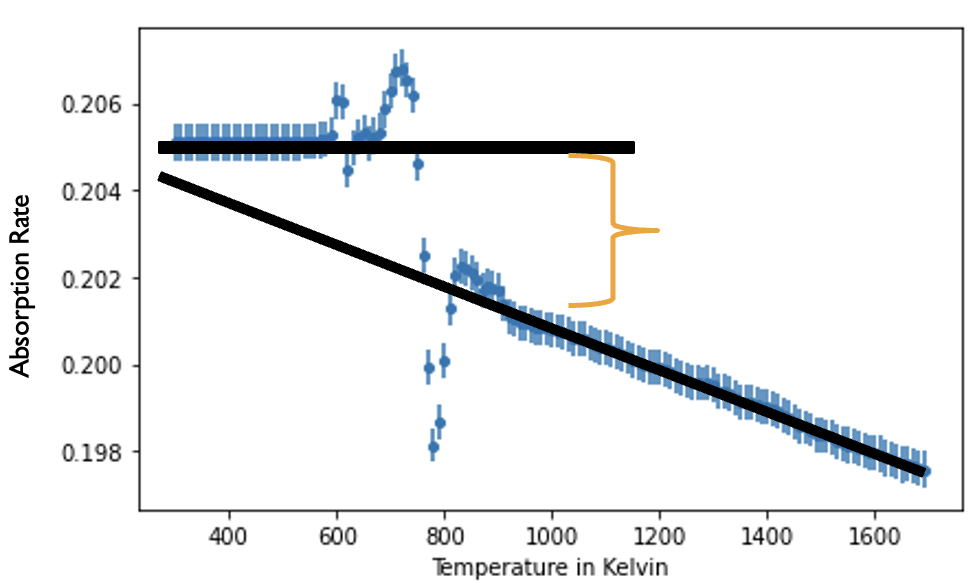}
    \caption{Temperature vs absorption rate over the second resonance, [18,23] eV, of $^{238}$U for homogeneous simple reflected source driven, 32\,eV source, geometry modeled in MCNP6.2 only of $^{238}$U and  $^{16}$O. The black lines are used to emphasize the shift in the data, where the lower line is the data after the shift, and the yellow brace is representative of the baseline defect.}
    \label{mcnpBaselineDefect}
\end{figure}

The significant implication of the work presented in this article is the identification of a fundamental flaw in the prediction of resonance escape probability (resonance absorption). For all resonances above the threshold, a small value of $\theta$ could result in an incorrect resonance escape probability, with potential errors of up to 1\% per resonance. This baseline defect is visually apparent in Figure~\ref{mcnpBaselineDefect}, as indicated by the yellow brace. The black lines in the plot highlight the trend in the data before and after the shift in physics. The baseline defect manifests as a noticeable shift in accuracy between these two black lines, representing the resonance escape probability with less accurate physics versus it with more accurate physics. This discernible shift in resonance escape probability illustrates the baseline defect that occurs with a low value of $\theta$.

\begin{figure}[H]
    \centering
    \includegraphics[width= 9cm, height=7cm]{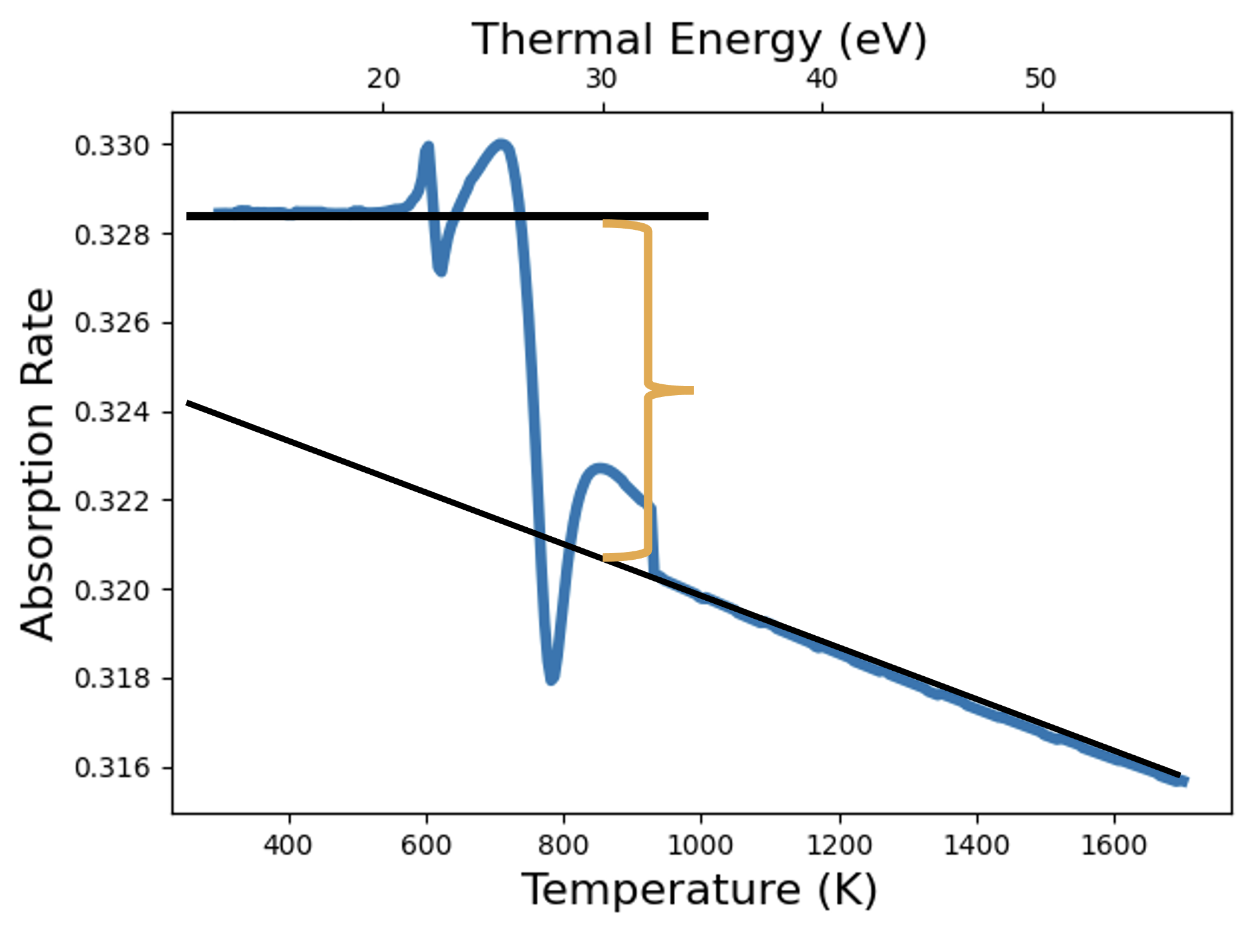}
    \caption{Temperature vs absorption rate over the second resonance, [18,23] eV, of $^{238}$U from a 0-dimensional deterministic source-driven transport code in written in Python. Neutrons are born at 32\,eV and terminated at 12\,eV. The geometry only includes $^{238}$U scattering and capture and $^{16}$O scattering. The red box highlights the second anomaly. The yellow brace is representative of the baseline defect.}
    \label{U8OBaselineDefectDeterministic}
\end{figure}

The baseline defect was also identifiable in the simple Python deterministic code. Figure~\ref{U8OBaselineDefectDeterministic} highlights the baseline defect present with the yellow brace.  

The final aspect of the baseline defect pertains to its directionality. Interestingly, the baseline defect does not exhibit a consistent direction. It has been observed to occur in either direction. Furthermore, the baseline defect is resonance and system-dependent, meaning that it could vary based on the specific isotopes present, the number density of materials, and the geometry of the problem. Consequently, the baseline defect may either diminish or intensify depending on these factors, highlighting the complexity of its manifestation.


\section{Applicability}

The area of applicability where the approximation does not work, and the anomalies are evident, is when there is a small length scale, measured in neutron mean free path, between fuel, material with absorbing resonances, and moderator, scattering material. This small length scale as shown earlier can be a heterogeneous problem geometry, if small enough, and any homogeneous geometry under these conditions. The first applicability is any MSR that can have the graphite porosity issue, like the MSRR discussed earlier. Another significant application to the nuclear industry is TRISO fuel. The fuel in TRISO particles is Uranium Oxycarbide (UCO) \cite{triso_p}. This is a ``homogeneous'' combination of uranium and carbon, like all of the above examples. In addition, TRISO particles also have an incredibly small length scale between fuel and graphite moderator. TRISO particles have a diameter of 0.08 to 0.1\,cm \cite{triso_p}, which is on the length scale where the anomalies have been identified. Further, the operating temperature of the gFHR is approximately 880-1100~K \cite{gFHR}. The anomaly is directly within this fuel temperature. 

\begin{figure}[H]
    \centering
    \includegraphics[width= 11cm, height=7cm]{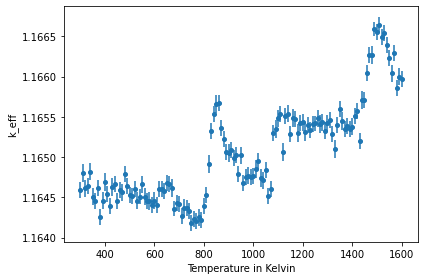}
    \caption{Temperature vs $k_{eff}$ for the Double Heterogeneous pebble case with reflective boundary conditions modeled in MCNP6.2.}
    \label{triso_hump}
\end{figure}

In order to see if the effect is present in this case, the specifications for the gFHR \cite{gFHR} pebble \cite{pebble_design} were modeled in the Double Heterogeneous style, such that the low-density graphite center, graphite outer-layer, and annular fuel were modeled discretely. However, the particles in the fuel annulus were modeled as a homogeneous fuel annulus. This pebble was modeled in isolation, and then reflective boundary conditions were applied to the outermost layer of graphite.  The results are shown in Figure~\ref{triso_hump}. Figure~\ref{triso_hump} is a plot of temperature vs $k_{eff}$ for the Double Heterogeneous pebble case described above with reflective boundary conditions modeled in MCNP6.2. In Figure~\ref{triso_hump} both anomalies can be seen. The first anomaly is clearly visible for the second and third resonances of $^{238}$U, at 850~K and 1491~K. In addition, the baseline defect can be seen clearly in this k-code calculation. Immediately after the switch in physics, at 850~K, there is a noticeable shift in the baseline of $k_{eff}$. The baseline defect in this case is on the order of 100\,pcm and is directly in the operating range of TRISO fuel. 

\section{Recommendations}

After thoroughly scrutinizing the anomalies caused by the approximation of $\theta$ being 400 in Equation~\ref{400kt}, we can now make recommendations on how to handle this effect. The recommendations that follow will all require the changing of $\theta$. Currently, it is not possible to change $\theta$ in any major Monte Carlo codes without modifying their source code. Therefore, the first recommendation is for all major Monte Carlo codes that use this approximation to make $\theta$ a user-definable variable. 

Once $\theta$ is a user-definable value, the authors would recommend that users complete a convergence study for the problem being modeled. A specific $\theta$ can not be universally prescribed because the $\theta$ that gives an accurate approximation will be problem dependent. Simulation time is expected to increase with $\theta$ because simulating target-in-motion physics is more computationally expensive than target-at-rest physics; the reason that the approximation was created in the first place. A convergence study will allow for the least expensive future simulations while ensuring that the baseline defect does not impact the output quantities of interest.

To demonstrate this, the source code of OpenMC \cite{openmc} was modified such that $\theta$ was a changeable value. The settings used when OpenMC performs the $\theta$=400 approximation are the following: \texttt{settings.temperature} \texttt{`method'} is set to \texttt{`interpolation'} and \texttt{`multipole'} is set to \texttt{`True'}, as well  \texttt{settings.resonance\_scattering} \texttt{`enable'} must be set to \texttt{`False'}. 

\begin{figure}[H]
    \centering
    \includegraphics[width=0.95\textwidth]{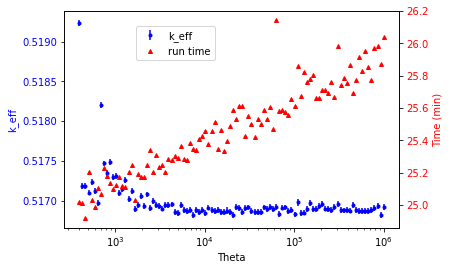}
    \caption{Theta($\theta$) vs $k_{eff}$ and Run Time with the x-axis plotted on a logarithmic scale. Modeled is the homogeneous version of the 0\% intrusion simple reflected pin-cell geometry at a constant temperature of 850~K in OpenMC with the source code modified to enable the changing of $\theta$. The following were all run on an AMD EPYC 7543P 32-core processor with 64 threads. }
    \label{changing_theta_time}
\end{figure}

Figure~\ref{changing_theta_time} is a plot of Theta($\theta$) vs $k_{eff}$ and Run Time with the x-axis plotted on a logarithmic scale in OpenMC using those settings along with the source code modified to enable the changing of $\theta$. All the runs were processed on an AMD EPYC 7543P 32-core processor with 64 threads. The geometry being modeled is the homogeneous version of the 0\% intrusion simple reflected pin-cell at a constant temperature of 850~K. It can be seen that time increases logarithmically as $\theta$ is increased. However, it can also be seen that for this problem $\theta$ does not need to be increased drastically to converge to the true value of $k_{eff}$. Additionally, in this problem, it only took an increase in run time by approximately 3\% to converge. This small increase in run time corrected a 200 pcm deficiency in $k_{eff}$.

\begin{figure}[H]
    \centering
    \includegraphics[width= 12.5cm, height=7cm]{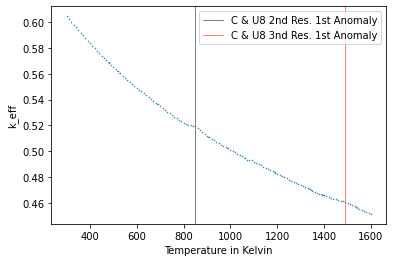}
    \caption{Temperature vs $k_{eff}$ at $\theta$=400. Modeled is the homogeneous version of the 0\% intrusion simple reflected pin-cell geometry in OpenMC with the source code modified to enable the changing of $\theta$.}
    \label{mv400}
\end{figure}

\begin{figure}[H]
    \centering
    \includegraphics[width= 12.5cm, height=7cm]{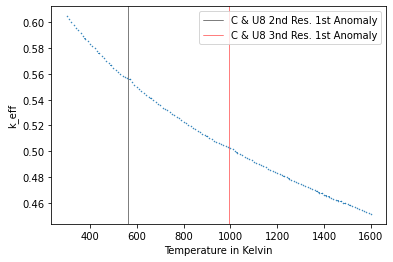}
    \caption{Temperature vs $k_{eff}$ at $\theta$=600. Modeled is the homogeneous version of the 0\% intrusion simple reflected pin-cell geometry in OpenMC with the source code modified to enable the changing of $\theta$.}
    \label{mv600}
\end{figure}

\begin{figure}[H]
    \centering
    \includegraphics[width= 12.5cm, height=7cm]{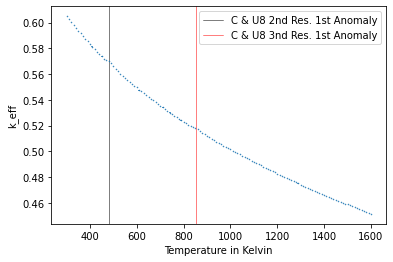}
    \caption{Temperature vs $k_{eff}$ at $\theta$=700. Modeled is the homogeneous version of the 0\% intrusion simple reflected pin-cell geometry in OpenMC with the source code modified to enable the changing of $\theta$.}
    \label{mv700}
\end{figure}

The other reason that a convergence study should be completed is that as $\theta$ is increased, the anomaly will move and the size will decrease. Figures~\ref{mv400}, \ref{mv600}, and \ref{mv700} show this effect. Figures~\ref{mv400}, \ref{mv600}, and \ref{mv700} were generated using the same OpenMC settings described above as well as with the source code modifications necessary to change $\theta$.  Additionally, it is important to note that OpenMC does account for the Doppler broadening of resonances and DBRC. Therefore, the following plots will have the correct slope in $k_{eff}$ for a system with changing temperature. The geometry being modeled is the homogeneous version of the 0\% intrusion simple reflected pin-cell, with different values $\theta$ for each. Figures~\ref{mv400}, \ref{mv600}, and \ref{mv700} are Temperature vs $k_{eff}$ at $\theta$=400, 600, and 700. As predicted, it can be seen that as $\theta$ is increased the anomalies seen move towards the left, and their size decreases. 

\section{Conclusion}

In conclusion, the approximation that target-at-rest is valid at for neutrons that are 400 times the energy of the target particle is not an accurate assumption for certain cases. Practical cases identified in this paper are TRISO fuel and when fuel seeps into the pores of the graphite in MSRs. The anomaly may lead to safety-relevant systematic errors of reactivity-temperature feedback coefficients. 
More generally, the approximation breaks down when there is a small length scale between fuel, a material with absorption resonances, and moderator, a scattering material.

Too small of a $\theta$, the current default setting of 400, value can result in a baseline defect that can lead to incorrect resonance escape probabilities that are off by 1\% per resonance. In addition, two separate anomalies were visually identified by changing temperature as a direct result of the switch between target-at-rest and target-in-motion physics. This article provides equations to predict the temperature-location of these anomalies given the system temperature and $\theta$. Additionally, this paper found that a TRISO fuel pebble is off by 100-200\,pcm at reactor temperature. The error in feedback coefficients was not analyzed, since it is design-specific.

The recommendations presented by the authors of this paper are as follows: 1) All major Monte Carlo codes that use this approximation make $\theta$ a user-defined variable. 2) Once this change is made, users should complete a convergence study to ensure that the $\theta$ value chosen is high enough to render the baseline defect and anomalies insignificant.

\newpage


\section*{Acknowledgments} 
This report was prepared by the research group of Dr. Sobes under award 31310021M0041 from Assistance Agreements, Nuclear Regulatory Commission. The statements, findings, conclusions, and recommendations are those of the author(s) and do not necessarily reflect the view of the Assistance Agreements or the US Nuclear Regulatory Commission.

This material is based upon work supported by the Department of Energy National Nuclear Security Administration through the Nuclear Science and Security Consortium under Award Number(s) DE-NA0003996.

This paper would like to acknowledge Emma Houston from the University of Tennessee-Knoxville for providing the Double Heterogeneous TRISO model in MCNP. 

\section{Declaration of Generative AI and AI-assisted technologies in the writing process}
During the preparation of this work, the authors used ChatGPT-3.5 to improve grammar and readability throughout the paper. After using this tool/service, the authors reviewed and edited the content as needed and take full responsibility for the content of the publication.
 \bibliographystyle{elsarticle-num} 
 \bibliography{editable}





\end{document}